\documentstyle[11pt,newpasp,epsf]{article}
\def\ts{\thinspace}
\begin{document}

\title{The Hosts of Ultraluminous Infrared Galaxies}

\author{D. B. Sanders}
\affil{Institute for Astronomy, University of Hawaii, 2680 Woodlawn Drive, 
       Honolulu, HI\ 96822:\ sanders@ifa.hawaii.edu}

\author{D.-C. Kim}
\affil{Institute of Astronomy and Astrophysics, Academia Sinica, P.O. 
       Box 1-87, Nankang, Taipei 11529, Taiwan:\ kim@asiaa.sinica.edu.tw}

\author{J. M. Mazzarella, and J. A. Surace}
\affil{IPAC, Caltech/JPL, MS 100-22, Pasadena, CA 91125:
\qquad\qquad\qquad mazz, jason@ipac.caltech.edu}

\author{J. B. Jensen}
\affil{Gemini Science Fellow, The Gemini North Observatory, 
670 N. A'ohoku Place, Hilo, HI 96720:\ jjensen@gemini.edu}

\begin{abstract}
Complete samples of ultraluminous infrared galaxies
(ULIGs\footnote{$L_{\rm ir} \equiv L(8-1000{\ts}\mu m) >
10^{12}{\ts}L_\odot$; unless otherwise stated, $H_{\rm o} =$
75{\ts}km{\ts}s$^{-1}$ Mpc$^{-1}$, $q_{\rm o} = 0$}) have been imaged at
R-band and K-band from Mauna Kea.  Here we present a preliminary analysis
of the host galaxy magnitudes and the 1-D radial profiles for a subset of
objects in the {\it IRAS} 1-Jy sample of ULIGs ($z${\ts}$<${\ts}0.3), and
compare these properties  with recently published data for ``low-$z$"
QSOs.  ULIGs in the 1-Jy sample reside in luminous hosts, with mean
luminosities $\sim${\ts}2.7{\ts}$L_{\rm K}^\ast$ ($\Delta_{\rm
K}${\ts}$\sim${\ts}0.7--11{\ts}$L_{\rm K}^\ast$), and
$\sim${\ts}2.2{\ts}$L_{\rm R}^\ast$ ($\Delta_{\rm
R}${\ts}$\sim${\ts}0.5--9{\ts}$L_{\rm R}^\ast$), values which are
remarkably similar in the mean and range for the hosts of low-$z$ QSOs.
Approximately one-third of ULIGs have single nuclei and radial profiles
that are closely approximated by a $r^{1/4}$-law over the inner
$\sim${\ts}2--10{\ts}kpc radius.  These ``E-like" hosts have half-light
radii, and surface brightness ($r_{1/2}$, $\mu_{1/2}$) similar to  QSO
hosts at R-band, but systematically smaller half-light radii than QSOs at
K-band.  
\end{abstract}



\section{Introduction}

ULIGs are an important class of extragalactic objects that appear to
powered by a mixture of starburst and AGN activity, both of which are
fueled by an enormous supply of molecular gas that has been funneled into
the nuclear region during the merger of two gas-rich spirals.  ULIGs rival
quasi-stellar objects (QSOs) in  bolometric luminosity, and there is
speculation that ULIGs may indeed represent an important stage in the
formation of QSOs as well as powerful radio galaxies, and that they may
also represent a primary stage in the formation of giant ellipticals (see
Sanders \& Mirabel 1996 for a more complete review).

A complete sample of 115 ULIGs ($f_{60} > 1${\ts}Jy) was compiled by Kim
(1995; see also Kim \& Sanders 1998) from redshift surveys of objects in
the {\it IRAS} Faint Source Catalog (FSC:  Moshir et al. 1992).  As the
nearest and brightest ULIGs, the ``1-Jy sample" provides the best list of
objects for detailed multiwavelength studies.  Optical spectroscopy for
the entire sample has recently been published by Veilleux, Kim, \& Sanders
(1999), and near-IR spectra have been published for 60\% of the total
sample by Veilleux, Sanders, \& Kim (1999).  High-resolution, tip-tilt,
optical and near-IR images have just recently been obtained for the entire
sample and these data are currently being reduced.

Here we present further analysis of the original data from Kim (1995) in
order to highlight the first intriguing results from a comparison of the
host galaxy properties of ULIGs with recently published data on the hosts
of QSOs.

\section{Imaging Observations -- The IRAS 1-Jy Sample}

Photometric CCD images of a subset of the 115 ULIGs in the IRAS 1-Jy
sample were originally obtained  at K-band\footnote{Kim (1995) observed
objects using a K$^\prime$ filter.  These data have been converted to the
standard K-band using the small correction terms given in Wainscoat \&
Cowie (1992)} (37/115) and R-band (83/115) using the QUIRC 256$\times$256
infrared camera (Hodapp, Rayner, \& Irwin 1992) and Tek 2048$\times$2048
optical CCD camera, respectively on the University of Hawaii (UH) 2.2m
telescope on Mauna Kea.  Details of the observations, data reduction, and
photometry can be found in Kim (1995).

\section{Host Galaxies of ULIGs}

One of the basic properties of ULIGs, and perhaps the most straightforward
to measure, is the total magnitude of the host galaxy and any point-like
nuclear source that may be present.  Kim (1995) found that the mean total
magnitudes of ULIGs in the 1-Jy sample were $\sim${\ts}3{\ts}$L_{\rm
K}^\ast$ and $\sim${\ts}2.7{\ts}$L_{\rm R}^\ast$, with a sample range of a
factor of $\sim${\ts}3$^{\pm 1}$ in both bands.  These mean values are
similar to the magnitudes of giant ellipticals (gEs) and brightest cluster
galaxies (BCGs).

\begin{figure}
\plotone{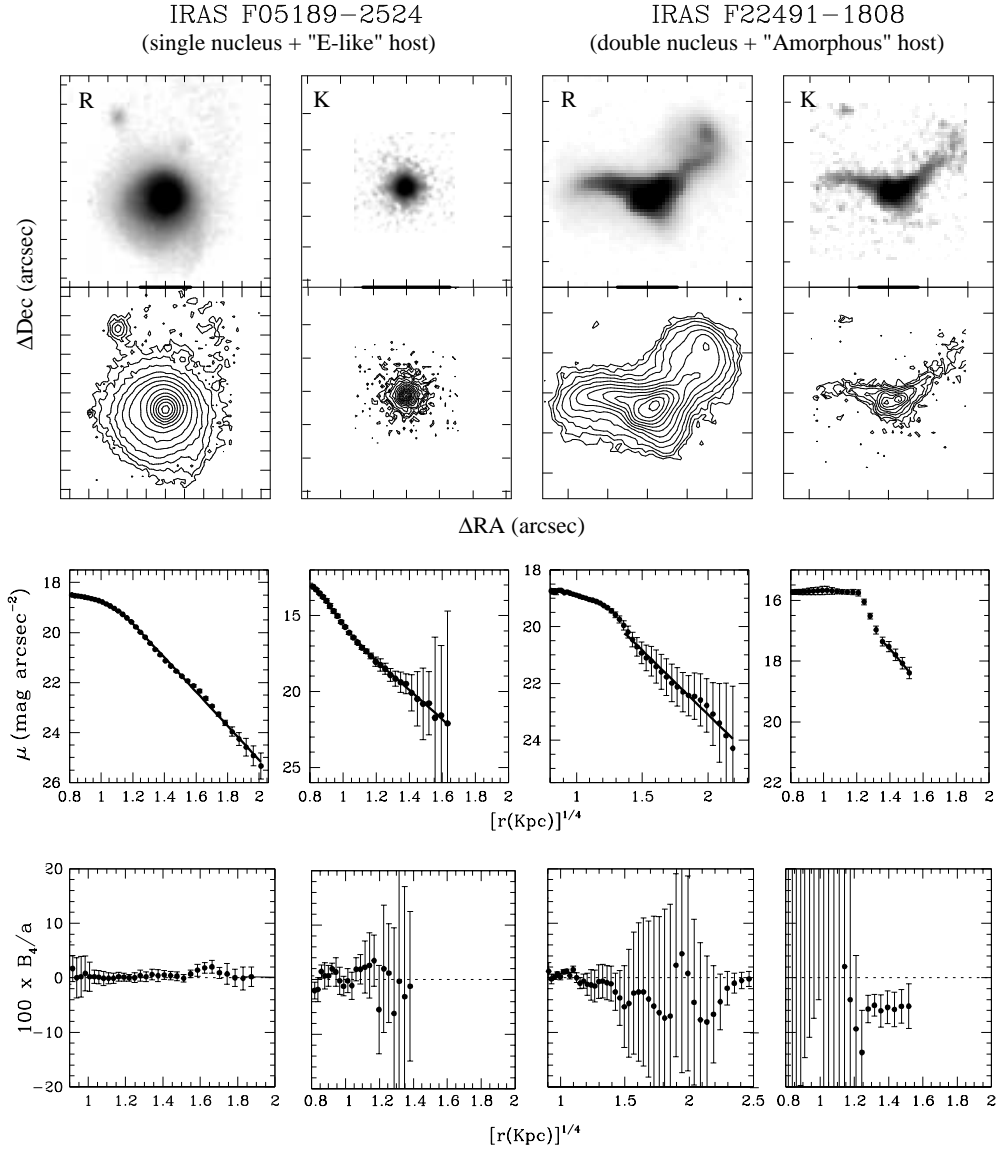}
\caption{R-band and K-band images and 1-D surface intensity profiles for
two ULIGs -- the ``IR-QSO" IRAS{\ts}F05189$-$2524 and the ``cool" ULIG
IRAS{\ts}22491$-$1808.  Rows 1 and 2 show the UH{\ts}2.2m tip-tilt images
in grayscale and contour form respectively.  North is to the top, east is
to the left, tick marks are at 10\arcsec\ intervals, and the solid bar
represents 10{\ts}kpc.  Rows 3 and 4 show the 1-D radial variation of
surface brightness and the parameter $B_4/a$ as a function of $r^{1/4}$.
The best $r^{1/4}$ fit to the surface brightness profile (excluding the
central 1.5\arcsec\ radius of the image) is given by the straight line.}
\end{figure}

Kim (1995) classified $\sim${\ts}1/3 of the R-band images of ULIGs as
elliptical-like based on a reasonable $r^{1/4}$-law fit to the radial
surface brightness profile over a relatively large range of host galaxy
radius, typically $\Delta r =${\ts}2--10{\ts}kpc.  An example of these
``E-like" ULIGs is IRAS{\ts}05189-2524 shown in Figure 1.  A majority of
these objects have Seyfert{\ts}1 optical or near-IR spectra, (similar to
the recent results reported by Zheng et al. 1999), hence the apt
description of these ``E-like" ULIGs as ``infrared QSOs".  An additional
$\sim${\ts}1/3 of the R-band images were classified as ``E/Sp" to describe
the fact that an $r^{1/4}$ law was a better fit than an exponential-disk
model over a large range of galactocentric radius.  The remaining
$\sim${\ts}1/3 of the 1-Jy ULIGs were divided between a few spiral-like
hosts (i.e. good exponential fit), highly distorted ``Amorphous" hosts
(typically closely spaced double nuclei where exponential-law and
$r^{1/4}$-law descriptions were equally bad fits; IRAS{\ts}22491-1808 in
Figure 1 is a good example), and a few objects with largely
non-overlapping disks where a single-host picture was clearly
inappropriate.

In comparing the R-band and K-band radial profiles of the ULIGs, there
were both widespread differences as well as similarities.  For example, in
the complete sample strong point-like nuclei at R-band are relatively
rare; the fraction of the total R-band light within the inner 2{\ts}kpc
radius is typically $<${\ts}20\%.  However, the K-band surface brightness
profiles are nearly always more centrally peaked, (e.g.
IRAS{\ts}05189-2524; see Figure 1).  As for the hosts, the distribution of
host types (``E-like", ``E/Sp", etc.) was generally the same at K-band as
that found at R-band with the only caveat being that the moderate depth of
the K-band images  meant that the galactocentric radius of the outermost
measured K-band contour was typically $\sim${\ts}2/3 that of the outer
R-band contour.

\section{Properties of ``E-like" ULIG Hosts}

\begin{figure} 
\plotone{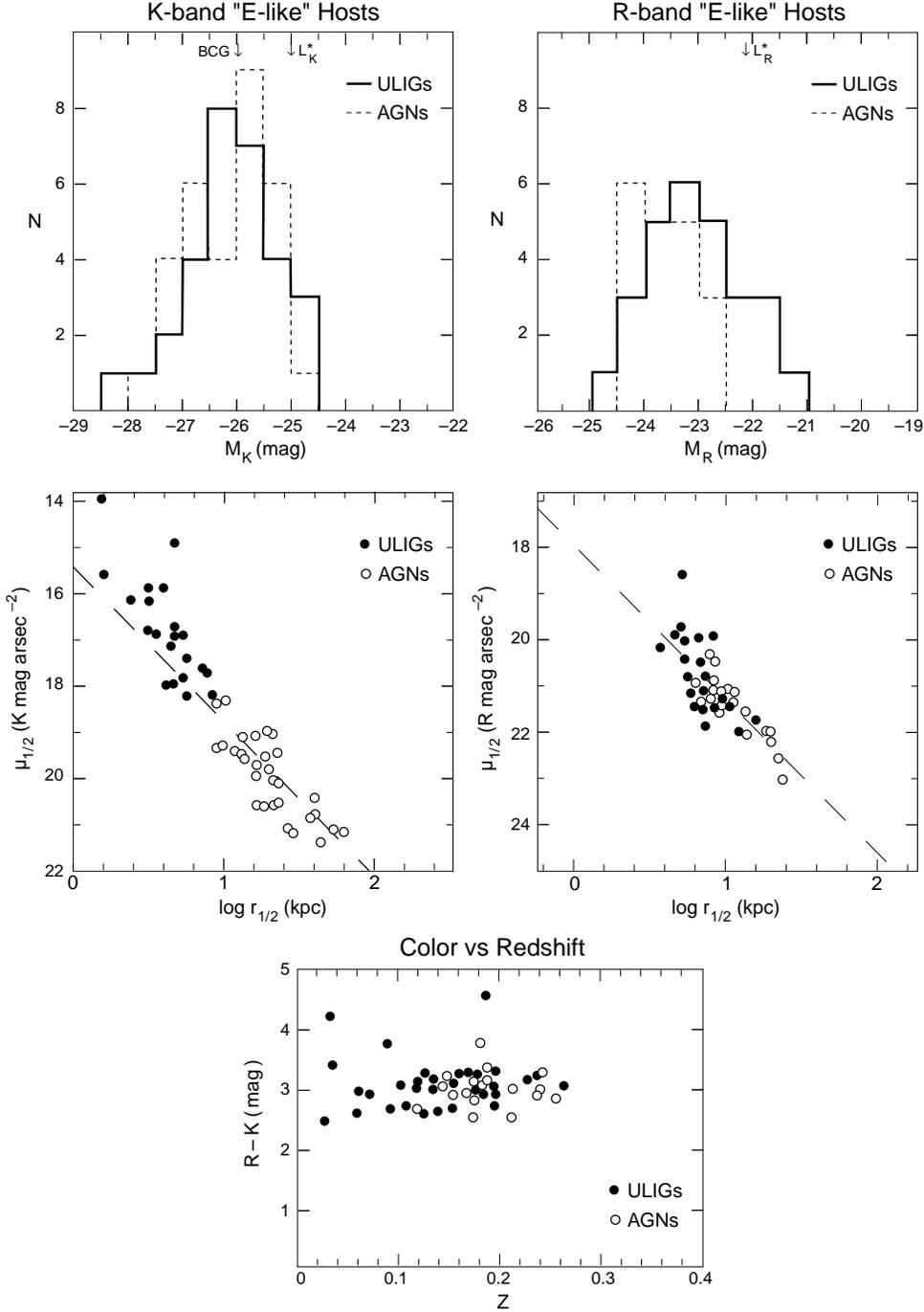} 
\vspace{-0.2in}
\caption{{\it [Note: For comparison with QSO data, all ULIG measured
parameters have been converted to H$_{\rm o} =${\ts}75
km{\ts}s$^{-1}$Mpc$^{-1}$, q$_{\rm o} =${\ts}0.]}\ R-band and K-band
magnitudes, effective radii and surface brightness ($r_{1/2}$,
$\mu_{1/2}$), and (R$-$K) colors for the hosts of those ULIGs with single
nuclei and $r^{1/4}$-like radial profiles (``E-like"), compared with
recently published K-band and R-band data for the hosts of AGNs. [Here we
use AGNs to refer to the composite K-band and R-band samples of ``low-$z$"
objects (i.e. RLQs$+$RQQs$+$RGs) as defined by Taylor et al. (1996) and
McLure et al.  (1999), respectively.]} 
\end{figure}

Figure 2 shows the distribution of host galaxy magnitudes for the
subsample of ``single-nucleus" objects in the IRAS 1-Jy sample, their
R$-$K colors and the half-light radii and surface brightness ($r_{1/2}$,
$\mu_{1/2}$) derived from an $r^{1/4}$-law fit to the 1-D radial
brightness profiles.  This analysis is similar to the analysis carried out
by Kim (1995), except here we have subtracted off any obvious nuclear
point source that may be present in the images of the 1-Jy ULIGs.  An
immediate result is that nearly all ULIGs reside in hosts with
luminosities above $L^\ast$.  The mean host luminosities are
2.7{\ts}$L_K^\ast$ and 2.2{\ts}$L_R^\ast$.  The radial surface brightness
profiles  over the range $\Delta r${\ts}$\sim${\ts}2--10{\ts}kpc are
well-fit by a deVaucouleurs $r^{1/4}$-law, whereas the light distribution
in the inner few kpc is more schizophrenic, most likely due to both heavy
dust obscuration as well as a recent luminous starburst.  Given these
observations, and the reasonable assumption from continuity arguments that
the non ``E-like" ULIGs will soon resemble their slightly more evolved
``E-like" cousins, it would appear that ULIGs in general reside in
luminous hosts that are similar in many ways to gEs and BCGs.

\subsection{Comparison with QSOs}

One of the most interesting new results from the current study of
``E-like" ULIGs is that their mean luminosities at R- and K-band are {\it
nearly identical} to that found previously for the hosts of QSOs (e.g.
McLeod \& Rieke 1994, 1995; Disney et al.  1995; Boyce et al.  1996;
Taylor et al. 1996; Sanders \& Surace 1997; Hutchings 1997; Bahcall et al.
1997; Surace 1998; McLure et al.  1999).  As shown in Figure 2, even the
range of luminosities for ULIG hosts appears similar to the range observed
for QSO hosts, typically 1--10{\ts}$L^\ast$ at both K-band (Taylor et al.
1996) and  R-band (McLure et al. 1999).  The range of R$-$K color of the
hosts of ULIGs and QSOs are similar (typically 2.5--3.5{\ts}mag), with the
exception of a few red ULIGs with R$-$K{\ts}$>${\ts}4{\ts}mag.  However,
the radial surface brightness distributions clearly show systematic
differences, with the half-light radii at R-band and K-band for ULIGs
being as much as a factor of 2.5 and 3.5 less respectively than found for
the hosts of QSOs.  Thus despite the overall similarity in host total
magnitudes, there is clearly an excess of emission in the inner disks of
IR-selected ULIGs as compared with optically-selected QSOs.

\section{Conclusions}

ULIGs in the local Universe ($z${\ts}$<${\ts}0.3) reside in luminous hosts
with mean luminosities $\sim${\ts}2--3{\ts}$L^\ast$ at R- and K-band.
Continuity arguments suggest that ULIG hosts are evolving into giant
ellipticals.  The properties of ULIGs which already have ``E-like" hosts
are quite similar to those of QSOs, but with an excess of emission at
K-band in the inner several kpc radius, perhaps due to a population of red
giants from an aging powerful circumnuclear starburst.  These results lend
further support to the hypothesis that ULIGs may be the infrared
precursors to optically-selected QSOs.

\acknowledgments
We thank Wendy Nakano and Karen Teramura for help in preparing the figures.

\end{document}